\documentclass[aps,prx,letterpaper,superscriptaddress,longbibliography, 11pt]{revtex4-2}

\usepackage{amsmath,amsthm,amsfonts}
\usepackage{multirow}
\usepackage{array}
\usepackage{soul}
\usepackage{float}
\usepackage{graphicx}
\usepackage{afterpage}
\usepackage{siunitx}
\usepackage{gensymb}

\usepackage[justification=raggedright,singlelinecheck=false,labelfont=bf,labelsep=quad, format=plain,font=small]{caption,subcaption}
\usepackage[colorlinks,citecolor=blue,urlcolor=blue,bookmarks=false,hypertexnames=true]{hyperref}
\usepackage{cleveref}

\newcommand{\sub}[1]{\textsubscript{#1}}
\newcommand{\degC}[1]{#1 \degree C}
\newcommand{\Ang}{\AA}

\begin{document}

\title{Atomic layer etching of niobium nitride using sequential exposures of O\sub2 and H\sub2/SF\sub6 plasmas}

\author{Azmain A. Hossain}
\affiliation{Division of Engineering and Applied Science, California Institute of Technology, CA 91125, USA}
\author{Sela Murphy}
\affiliation{Department of Material Science and Engineering, University of California Berkeley, CA 94720, USA}
\author{David Catherall}
\affiliation{Division of Engineering and Applied Science, California Institute of Technology, CA 91125, USA}
\author{Anthony J. Ardizzi}
\affiliation{Division of Engineering and Applied Science, California Institute of Technology, CA 91125, USA}
\author{Austin J. Minnich}
\email{aminnich@caltech.edu}
\affiliation{Division of Engineering and Applied Science, California Institute of Technology, CA 91125, USA}
\date{\today}

\begin{abstract}

Niobium nitride (NbN) is a metallic superconductor that is widely used for superconducting electronics due to its high transition temperature ($T_c$) and kinetic inductance. Processing-induced damage negatively affects the performance of these devices by mechanisms such as microwave surface loss. Atomic layer etching (ALE), with its ability to etch with Angstrom-scale control and low damage, has the potential to address these issues, but no ALE process is known for NbN. Here, we report such a process consisting of sequential exposures of O\sub{2} plasma and H\sub2/SF\sub6 plasma. Exposure to O\sub{2} plasma rather than O\sub{2} gas yields a greater fraction of Nb in the +5 oxidation state, which is then volatilized by NbF\sub{5} formation with exposure to an H\sub2/SF\sub6 plasma. The SF\sub{6}:H\sub{2} flow rate ratio is chosen to produce selective etching of Nb\sub{2}O\sub{5} over NbN, enabling self-limiting etching within a cycle. An etch rate of 1.77 \Ang/cycle was measured at \degC{125} using \textit{ex situ} ellipsometry. The $T_c$ of the ALE-etched film is higher than that of an RIE-etched film of a similar thickness, highlighting the low-damage nature of the process. These findings have relevance for applications of NbN in single-photon detectors and superconducting microresonators.

\end{abstract}

\maketitle

\newpage

\section{Introduction}

Niobium nitride (NbN) is a material of interest for superconducting electronics due to its high superconducting critical temperature $T_c$ (up to 16 K) \cite{Shy1973Dec, Kang2011Feb}, kinetic inductance ($>$ 150 pH/$\Box$) \cite{Niepce2019Apr, Yoshida1992Dec, Anferov2020Feb}, and high absorption coefficient in the infrared and optical wavelengths \cite{Semenov2009Aug, Anant2008Jul, Verevkin2002Jun}. These properties make it promising for devices such as kinetic inductance detectors (KIDs) \cite{Ariyoshi2013May, Mazzocchi2018}, superconducting nanowire single-photon detectors (SNSPDs) \cite{Goltsman2001Aug, Korneev2004Jun}, qubits \cite{Yu2002May, Kim2021Sep}, hot electron bolometers (HEBs) \cite{Guillet2008Apr, Kooi2007Feb}, and other superconducting quantum devices \cite{deGraaf2018Jun, Iosad2002Aug, Annunziata2010Oct, Zhang2019Aug, Niepce2019Apr, Anferov2020Feb}. The high $T_c$ of NbN film compared to other common superconductors such as Al and TiN (1.2 K and 5.5 K, respectively) allows for higher operating temperatures which alleviates space, weight, and power constraints of space missions \cite{Baselmans2022Sep, Ulbricht2021Mar, Mazin2020Apr}. 

State-of-the-art NbN devices are typically fabricated using reactive ion etching (RIE) with fluorine-containing plasmas \cite{Hu2020Nov, Guo2020Jun, EsmaeilZadeh2020Jul, Knehr2019Oct}. RIE fabricated devices such as superconducting microwave resonators based on NbN have been reported with single-photon internal quality factors $Q_i > 10^5$ \cite{Frasca2023Oct, Carter2019Aug, Schmieden2024Dec}. NbN SNSPDs have been demonstrated with $>95 \%$ system efficiency and timing jitter $< 70$ ps \cite{Zhang2019Oct, Zhang2017Dec, Hu2020Nov}. However, the figures of merit of these devices are negatively impacted by fabrication-induced damage. For example, the quality factor of superconducting microresonators is presently thought to be limited by microwave surface loss associated with two-level systems (TLS) at various interfaces \cite{Gao2007Mar, Barends2008Jun, Gao2008Apr}. In addition, the constriction factor of SNSPDs is often limited by sidewall roughness caused in part by RIE-induced damage, which will in turn reduce the cutoff operational wavelength and limit internal efficiency near the long wavelength cutoff \cite{Hadfield2007Dec, Kerman2007Mar, Colangelo2022Jul, Craiciu2023Feb, Ilin2008Feb, Frasca2019Aug}.


Typical dry etching processes for pattern transfer employ plasmas with energetic ions, making sample damage difficult to avoid for non-negligible etch rates. Atomic layer etching (ALE) is an emerging nanofabrication process with potential to overcome these limitations \cite{Lill2016, George2020Jun, Sang2020}. The first ALE processes were plasma-based directional methods based on sputtering a modified surface layer with a reduced sputtering threshold using low-energy ions \cite{Sakaue1990Nov, Horiike1990May}. Recently, thermal ALE processes have been reported which rely on sequential, self-limiting surface chemical reactions without the requirement of plasma exposures \cite{George2016}. In thermal ALE, the film surface is modified to form a non-volatile layer that can then be removed in various ways. The first report of thermal ALE used ligand-exchange transmetalation reactions \cite{Lee2015May}. Since then, other approaches have been reported, including conversion-etch processes \cite{DuMont2017Mar, Abdulagatov2020Feb}, temperature cycling \cite{Miyoshi2017Apr, Song2018Sep}, and others \cite{Ishii2017May, George2020Jun}. Plasma-thermal ALE has also been reported in which radical species from the plasma are used for modification or volatilization \cite{Fischer2023Aug, Chittock2023Aug, Catherall2024Sep, Wang2023May, Chen2024Oct}. Plasma and thermal ALE processes have now been reported for various dielectrics, metals and semiconductors including Al\sub{2}O\sub{3} \cite{Zywotko2018Nov, Lee2015May}, SiO\sub{2} \cite{DuMont2017Mar, Rahman2018Sep, Catherall2024Sep}, InP \cite{Park2006Jul, Ko1993Nov}, GaAs \cite{Ko1993Nov, Aoyagi1992Feb}, Cu \cite{Mohimi2018Aug, Sheil2021Jan}, W \cite{Johnson2017Oct, Xie2018Mar, Xie2020Jan}, and others \cite{Sang2020, Fang2018Dec, Fischer2021May, Saare2023Jun, Pina2024Aug}. Surface smoothing of etched surfaces using isotropic ALE has also been reported for various metals and semiconductors \cite{Zywotko2018Nov, Kanarik2017Sep, Ohba2017May, Gerritsen2022Dec, Konh2022Dec}.

Although a number of ALE processes have been reported for materials of relevance to the semiconductor industry \cite{Kanarik2015, Oehrlein2015Rev3}, fewer studies have focused on materials relevant to quantum devices \cite{Hossain2023Sep, Chen2024Oct}. At present, no ALE process has been reported for NbN. Several ALE processes have been demonstrated for a related material, TiN, based on leveraging the volatility of higher oxidation state fluorides compared to lower oxidation state fluorides \cite{Lee2017TiN, Hossain2023Sep}. However, due to the complexity of the oxidation chemistry of Nb, whether these approaches are applicable to NbN is unclear.

Here, we report a plasma-thermal ALE process for NbN using sequential exposures of O\sub{2} plasma and SF\sub{6}/H\sub{2} plasma. The O\sub{2} plasma is used to preferentially form the +5 oxidation state of Nb over the lower oxidation states, which then facilitates volatilization as NbF\sub{5} by exposure to a H\sub{2}/SF\sub{6} plasma mixture. The etch rate is measured to be 1.77 \Ang/cycle at \degC{125} using \textit{ex situ} ellipsometry. The etched surface was found to exhibit a 55\% decrease in surface roughness and a 59\% decrease in surface oxygen concentration over 50 cycles. The superconducting transition temperature of the ALE-etched sample is higher than that of an RIE-etched film of a similar thickness, indicating that ALE induces less damage in the film.  Our findings suggest that our NbN ALE process could be used to mitigate surface loss and hence improve the performance of NbN superconducting electronics.

\begin{figure*}
\centering
\includegraphics[width=\textwidth]{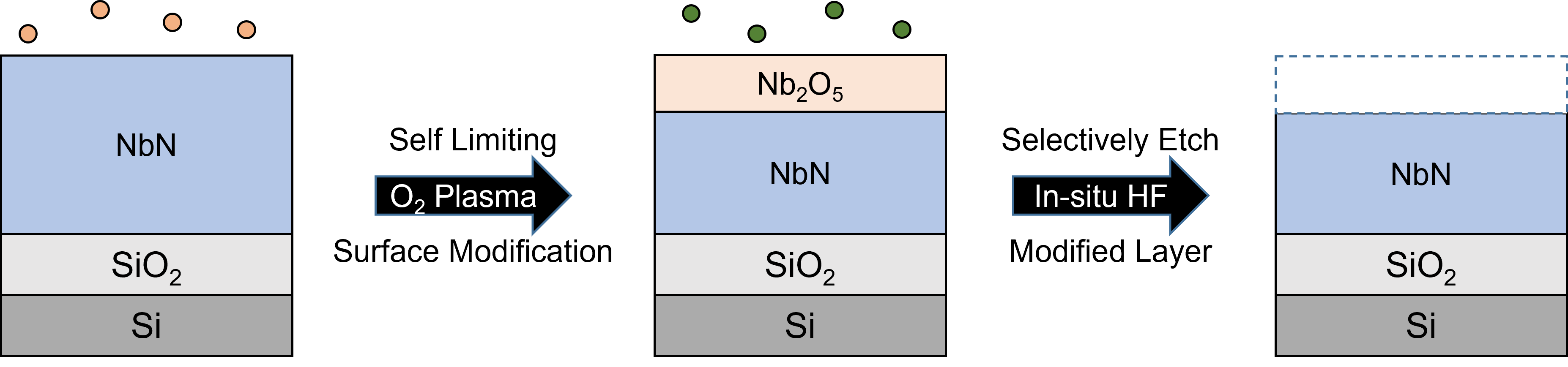}

\caption{Schematic of the NbN ALE process. The surface is first modified by exposure to O\sub{2} plasma (orange dots) to oxidize NbN to its +5 oxide of Nb\sub{2}O\sub{5}. Nb\sub{2}O\sub{5} is then exposed to an SF\sub6/H\sub2 plasma (green dots) to produce volatile etch products.}
\label{fig:NbNSchematic}
\end{figure*}

\section{Methods} \label{subsec:expmetd}

ALE processes were performed in an Oxford Instruments FlexAL II atomic layer deposition (ALD) system with an inductively coupled plasma source \cite{vanHemmen2007May}. The temperature of the substrate table was maintained at \degC{125} as measured by the substrate table thermometer. Before loading the sample into the chamber, preconditioning was performed to keep the chamber conditions as consistent as possible between different runs on the shared tool. The preconditioning consisted of 10 minutes of an SF\sub6/O\sub2 plasma to clean the chamber of contaminants with volatile fluoride species, followed by 300 cycles ($\sim$ 50 nm) of alumina ALD \cite{vanHemmen2007May}, and another 5 minutes of SF\sub6/O\sub2 plasma to react with the remaining precursor. Alumina was used as it forms a stable AlF\sub3 layer when exposed to SF\sub6 plasma. 

The plasma-thermal ALE process of this work is illustrated in \Cref{fig:NbNSchematic}. Each ALE cycle started with 6 s exposure of oxygen plasma to oxidize the surface of NbN to Nb\sub{2}O\sub{5}, followed by a 15 s purge. Then, H\sub{2} and SF\sub{6} gas were flowed at a fixed flow rate ratio before striking a plasma and exposing for 30 s. After another 15 s purge, the cycle is complete. For each plasma step, the chamber pressure is held at 100 mTorr and the plasma is struck at 400 W. After ALE processing, the chamber is pumped down for 60 s before unloading the sample. The overall process has a cycle time of $\sim$ 68 s/cycle.

The etched amounts were calculated based on thickness measurements of a $10\times10$ mm$^2$ chip before and after processing using \textit{ex situ} ellipsometry on a J.A. Woolam M2000 ellipsometer. The film thickness was determined by scanning 9 points at $60^\circ$ and $70^\circ$ from 370 - 1000 nm. NbN optical data was obtained from Ref.~\cite{Medeiros2019Jul}. The data was fit to an NbN/SiO\sub2/Si film stack to determine the thickness at each point. The thickness and uncertainty reported are the average and standard deviation of the 9 ellipsometry data points, respectively.

The film surface topography was characterized using a Bruker Dimension Icon atomic force microscope (AFM) over a $250 \times 250$ nm$^2$ area. The raw height maps collected on the AFM were processed by removing tilt using the Bruker Analysis linear flatten function. The surface roughness and power spectral density (PSD) were computed from the plane-fit height maps using procedures outlined in previous literature \cite{Gerritsen2022Dec, Catherall2024Sep, Hossain2023Sep}. The PSD provides a quantitative measure of the lateral distance over which the surface profile varies in terms of spatial frequencies \cite{Elson1995Jan, Jacobs2017Jan}.

Film composition was characterized using X-ray photoelectron spectroscopy (XPS) performed on a Kratos Axis Ultra X-ray photoelectron spectrometer using a monochromatic Al K$\alpha$ source. Depth profiling was performed using an Ar ion beam with a 25 s interval for each cycle. The estimated milling depth was calculated based on initial and final film thickness measured by \textit{ex situ} ellipsometry and assuming a constant ion milling rate. The XPS data was analyzed in CASA-XPS from Casa Software Ltd using bond binding energies and fitting procedures outlined in prior literature \cite{Yang2018May, Lubenchenko2018Jul, Darlinski1987Jun, Jouve1996Oct}. 

Cryogenic electrical measurements were performed on a Quantum Design DynaCool Physical Property Measurement System (PPMS). The NbN films were connected to the PPMS sample holder by four aluminum wires, wirebonded with a Westbond 7476D Wire Bonder. Film resistivity $(\rho)$ was measured using a 4-point probe setup. The resistivity was measured from 5 K to 20 K, from which the superconducting critical temperature ($T_c$) of the films was computed. The $T_c$ reported is defined as the temperature value at which resistivity drops to half of the 20 K resistivity value, and the uncertainty reported is half the width of the superconducting transition temperature range $(\Delta T_c)$.

NbN films on a Si wafer were provided by the Jet Propulsion Laboratory. The films were deposited by reactive sputtering at room temperature on a high resistivity $500$ \textmu m Si (100) wafer with a $240$ nm thermally grown SiO\sub{2} buffer layer. The initial thickness was $30 \pm 1$ nm. The process followed those developed for NbN SNSPDs \cite{Guillet2008Apr, Dane2017Sep, Luo2023Aug, Stepanov2024Feb}. Individual sample chips were obtained by dicing the 4 inch wafer into $10\times10$ mm$^2$ squares.

The resistivity at 20 K and $T_c$ of the films was measured as $291\ \mu\Omega$cm and $11.22\ \pm\ 0.14$ K, respectively. The $T_c$ of our samples is within $10\%$ of those of sputtered NbN films on Si reported in literature \cite{Zhang2019Oct, Stepanov2024Feb}. Several sets of films were prepared for various portions of the study. To develop the ALE process, two samples were oxidized with either O\sub{2} gas (120 s at 100 mTorr) or O\sub2 plasma (120 s at 100 mTorr and 400 W) at \degC{125}. Both samples were etched with SF\sub{6} plasma at (15 s at 100 mTorr and 400 W) at \degC{125} prior to oxidation. The pre-processing removed the NbN native oxide and also etched $\sim3$ nm of NbN to emulate oxidation in the middle of an ALE process.

To test etch selectivity, niobium oxide films were prepared. NbN samples were exposed to an O\sub{2} plasma (120 s at 100 mTorr) at \degC{125}. The oxidized sample was then exposed to an H\sub{6}/SF\sub{6} plasma (30 s at 100 mTorr) at \degC{125} for varying gas ratios. 

NbN films were also etched using RIE as a reference for surface roughness and electrical properties. The RIE-etched films were prepared in an Oxford Instruments Plasmalab 100 ICP-RIE 380 System. NbN films were exposed to an SF\sub{6}/Ar plasma (20 s at 10 mTorr and 200 W ICP with 100 W Bias) at \degC{20}. Plasma parameters were based on reported fluorine-based RIE parameters for the fabrication of NbN SNSPDs \cite{EsmaeilZadeh2020Jul, Knehr2019Oct, Guo2020Jun, Henrich2012Oct}. The RIE-treated films were etched from 30.1 nm to 22.1 nm for a total etched thickness of $8.0 \pm 0.3$ nm. For comparison, 50 cycle ALE-treated films were etched from 30.0 nm to 21.1 nm for an etched thickness of $8.9 \pm 0.1$ nm.

\section{Results}

\subsection{NbN oxidation using O\sub{2} plasma}\label{subsec:oxyxps}

Our NbN ALE process was inspired by our prior work on TiN ALE using sequential O\sub{2} gas and H\sub{2}/SF\sub{6} plasma mixture exposures \cite{Hossain2023Sep}. As a first attempt, we used the same sequence on NbN. However, unlike for TiN, we observed that the etch rate plateaued towards zero after a few tens of cycles. This outcome is not unexpected because the oxidation chemistry of Nb is notoriously complex, with Nb taking on oxidation states ranging from +2 to +5 \cite{Greenwood1997Dec}. NbF\sub{5} is the most volatile fluoride, and its formation is necessary for etching \cite{Fairbrother1951Jan, Ehrlich1955Dec, Gortsema1965Feb}, meaning that the desired surface oxide is Nb\sub{2}O\sub{5}. 

Considering these points, we hypothesized that the plateauing etch rate could be attributed to the formation of lower oxidation state fluorides with low volatility. A potential route to circumvent this issue is oxidation via oxygen plasma. Oxygen plasma has been reported to oxidize with lower activation energy compared to oxidation by molecular oxygen \cite{Dias2023Aug}. There is also previous literature on the oxidation of niobium metal showing that oxygen plasma produces an oxide with a higher concentration of Nb\sub{2}O\sub{5} relative to the native oxide \cite{Zhang2019May, Marks1983Mar, Giaccone2022Oct, Hu1989Sep}.

\begin{figure}
\centering
{\includegraphics[width = 0.7\textwidth]{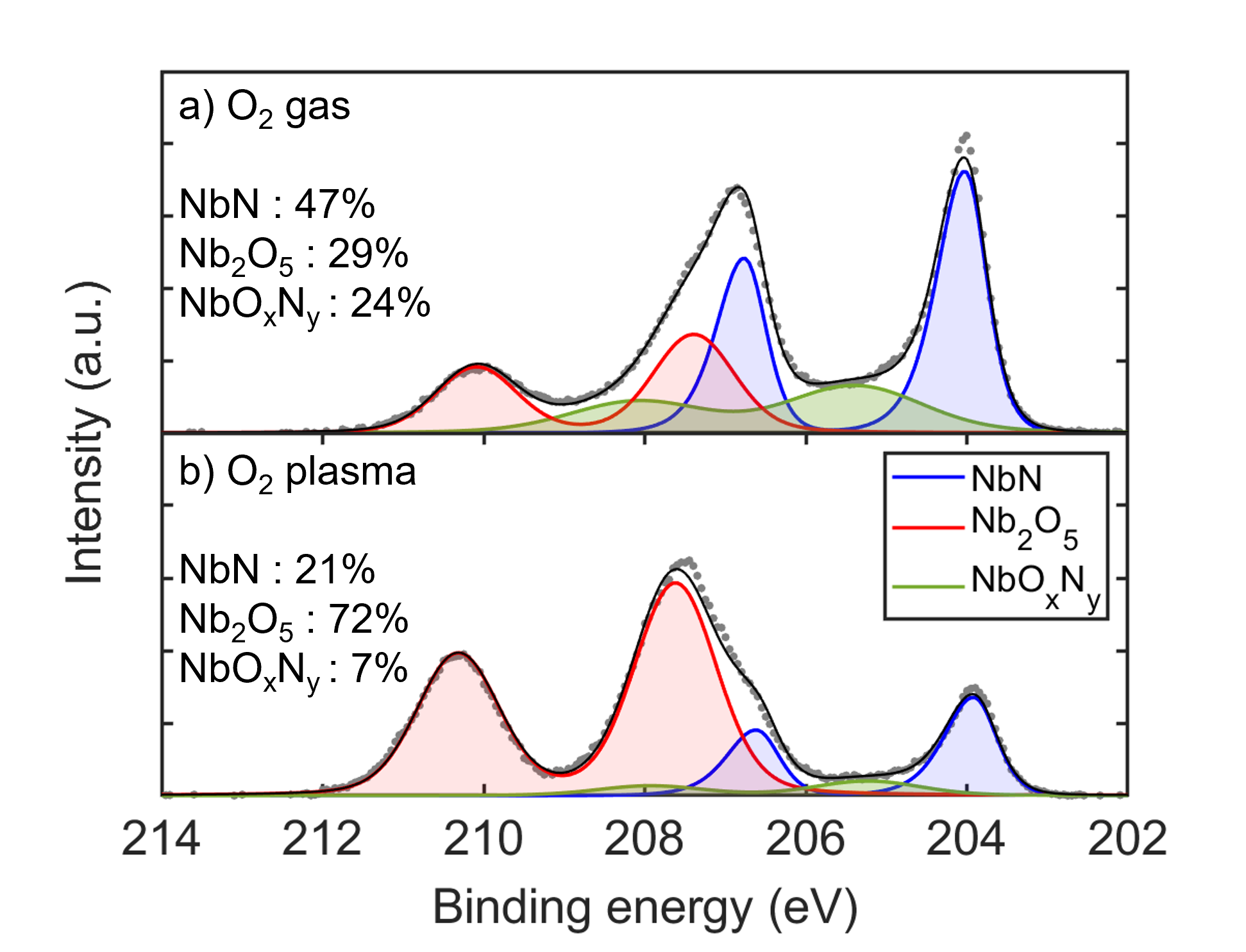}
\phantomsubcaption\label{fig:O2gas}
\phantomsubcaption\label{fig:O2plasma}
}

\caption{Surface XPS spectra showing Nb 3d spectra for NbN films exposed to (a) O\sub2 gas and (b) O\sub2 plasma for 120 s. The measured (gray dots) and fit spectra (black lines) intensity are reported in arbitrary units (a.u.) against the binding energy on the x-axis. The y-axis scales are identical for both spectra. The percent concentrations of the NbN, Nb\sub{2}O\sub{5} and NbO\sub{x}N\sub{y} bonds are reported in the figure.}
\label{fig:OxySurf}
\end{figure}

To verify that oxygen plasma exposure would produce a higher fraction of Nb\sub{2}O\sub{5}, we prepared two samples oxidized with either O\sub2 gas or plasma, as described in \Cref{subsec:expmetd}, and characterized the surface chemical environment with \textit{ex situ} XPS. \Cref{fig:OxySurf} shows the Nb 3d spectra for each sample. We observe three peaks, each consisting of a doublet due to spin-orbit coupling. These peaks are assigned as NbN, NbO\sub{x}N\sub{y}, and Nb\sub2O\sub5 \cite{Yang2018May, Lubenchenko2018Jul, Darlinski1987Jun, Jouve1996Oct}. The binding energies are listed in \Cref{subsec:regxps}. The Nb\sub{2}O\sub{5} subpeaks correspond to the +5 oxidation state of Nb when bonded to oxygen. Between the +3 oxidation state (corresponding to NbN) and +5 oxidation state, another component is observed. However, the composition of this component is difficult to determine as it can correspond to oxides of lower oxidation state such as NbO\sub{2} or an oxynitride such as NbON both of which have very similar binding energies \cite{Altoe2022Apr, Yang2018May, Lubenchenko2018Jul, Darlinski1987Jun, Jouve1996Oct}. In our analysis we denote the middle subpeaks by NbO\sub{x}N\sub{y} as it is likely a combination of NbO\sub{2} and NbON. 

The presence of Nb\sub{2}O\sub{5} can be quantified by computing the area fraction for each component for each sample. The results of the analysis are reported on \Cref{fig:OxySurf}. After exposure to O\sub2 gas, \Cref{fig:O2gas} shows almost equal fractions of Nb\sub{2}O\sub{5} and NbO\sub{x}N\sub{y} with 29\% Nb\sub{2}O\sub{5} bonding. \Cref{fig:O2plasma} shows the surface spectra after exposure to O\sub2 plasma, with a Nb\sub{2}O\sub{5} fraction of 72\%. These results confirm that Nb oxidation using a plasma forms more of the +5 oxidation state oxide and less of the lower oxidation state species. The +5 oxidation state is preferred as it forms volatile fluorides which lead to etching. The use of oxygen plasma is therefore advantageous over the oxygen gas for facilitating the etching of NbN.


\subsection{Selectivity of SF\sub{6}/H\sub{2} plasma} \label{subsec:sf6h2}

\begin{figure}
\centering
{\includegraphics[width = \textwidth]{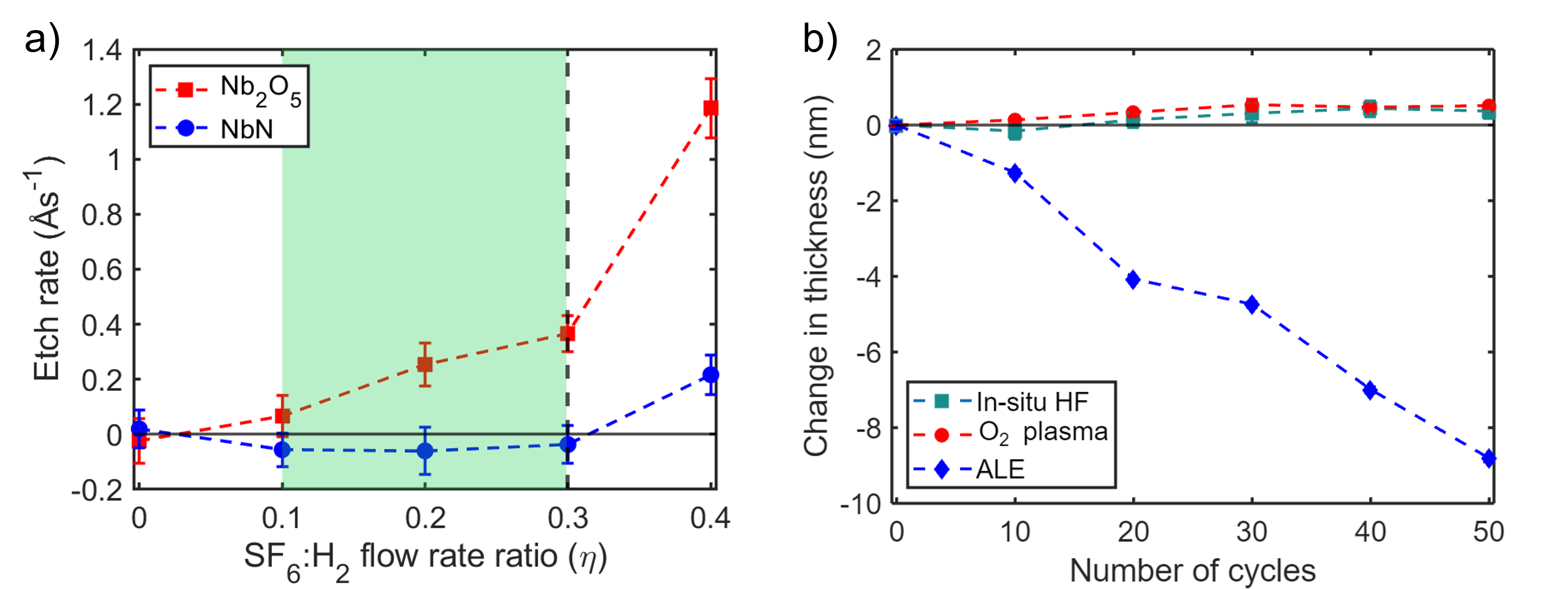}
\phantomsubcaption\label{fig:insituHF}
\phantomsubcaption\label{fig:thicknesscycle}
}
\caption{(a) Etch rate of NbO\sub{x} and NbN versus the SF\sub{6}:H\sub{2} flow rate ratio. The green shaded area represents the flow rate ratios for which selective etching of NbO\sub{x} over NbN was achieved. The black dashed line at a ratio of 0.3 represents the ratio used in the ALE experiments, and a solid line at 0 \Ang/cycle is plotted as a reference. (b) NbN thickness change versus  number of cycles with exposure only to O\sub2 plasma (red circles), \textit{in situ} HF (green squares) and full ALE process at \degC{125} (blue diamonds). A solid line is plotted at 0 nm change in thickness as a reference. The dashed lines are guides to the eye.}
\end{figure}

We next examine the etch rates of Nb\sub2O\sub5 and NbN films for varying SF\sub{6}:H\sub{2} flow rate ratio, $\eta$. The use of H\sub2/SF\sub6 plasma was motivated by noting that HF weakly fluorinates TiN without increasing the oxidation state \cite{Lee2017TiN}, but fluorine radicals cause spontaneous etching of NbN \cite{Harada1981Jan, Meng1989Mar}. Due to practical limitations, we are unable to use HF vapor, so we used an H\sub{2}/SF\sub{6} plasma mixture instead. H\sub{2}/SF\sub{6} plasma mixtures have long been known to produce vibrationally-excited HF which has been exploited for realization of chemical lasers \cite{Obara1974Apr, Hinchen1970Nov}. Prior studies using a hydrogen and fluorine-containing gas mixture plasma have shown that the products formed depend on the gas flow rate ratio. In particular, the formation of HF at sufficiently high H\sub{2}:SF\sub{6} flow rate ratio has been studied as it leads to etch selectivity \cite{Pankratiev2020DecHF, Volynets2020HF, Jung2020Mar, Gil2023Jul, Miyoshi2021Dec, Catherall2024Sep, Hossain2023Sep}.

\Cref{fig:insituHF} shows the etch rates of NbN and Nb\sub2O\sub5 versus $\eta$ at \degC{125}. For $\eta \lesssim 0.1$, negligible etching is observed in both films. For $\eta \geq 0.1$, we observe spontaneous etching of Nb\sub{2}O\sub{5}, with the etch rate increasing with $\eta$. For NbN, we observe no etching for $\eta \leq 0.3$. Instead, negative etch rates $\sim -0.05$ \Ang s$^{-1}$ are observed corresponding to an increase in the thickness of the film, which we assume to be growth of non-volatile lower oxidation state niobium fluorides. For $\eta \geq 0.4$, etching of NbN is observed. The etch selectivity of the oxide over the nitride may be attributed to the \textit{in situ} formation of vibrationally-excited HF along with negligible fluorine radical concentration for $0.1 < \eta \leq 0.3$, as has been reported previously \cite{Pankratiev2020DecHF, Volynets2020HF, Jung2020Mar, Hossain2023Sep, Gil2023Jul}. For $\eta \geq 0.4$, the concentration of F radicals becomes sufficient to spontaneously etch the NbN, leading to etching for both films. From our measurements, we find that $0.1 \leq \eta \leq 0.3$ achieves selective etching of Nb\sub2O\sub5 over NbN. To obtain the highest etch selectivity of Nb\sub2O\sub5 over NbN, we select $\eta = 0.3$ for our experiments. This ratio is used throughout the rest of the paper and referred to as \textit{in situ} HF.

\subsection{NbN ALE using O\sub2 and \textit{in situ} HF plasmas}

\begin{figure}
\centering
{\includegraphics[width = \textwidth]{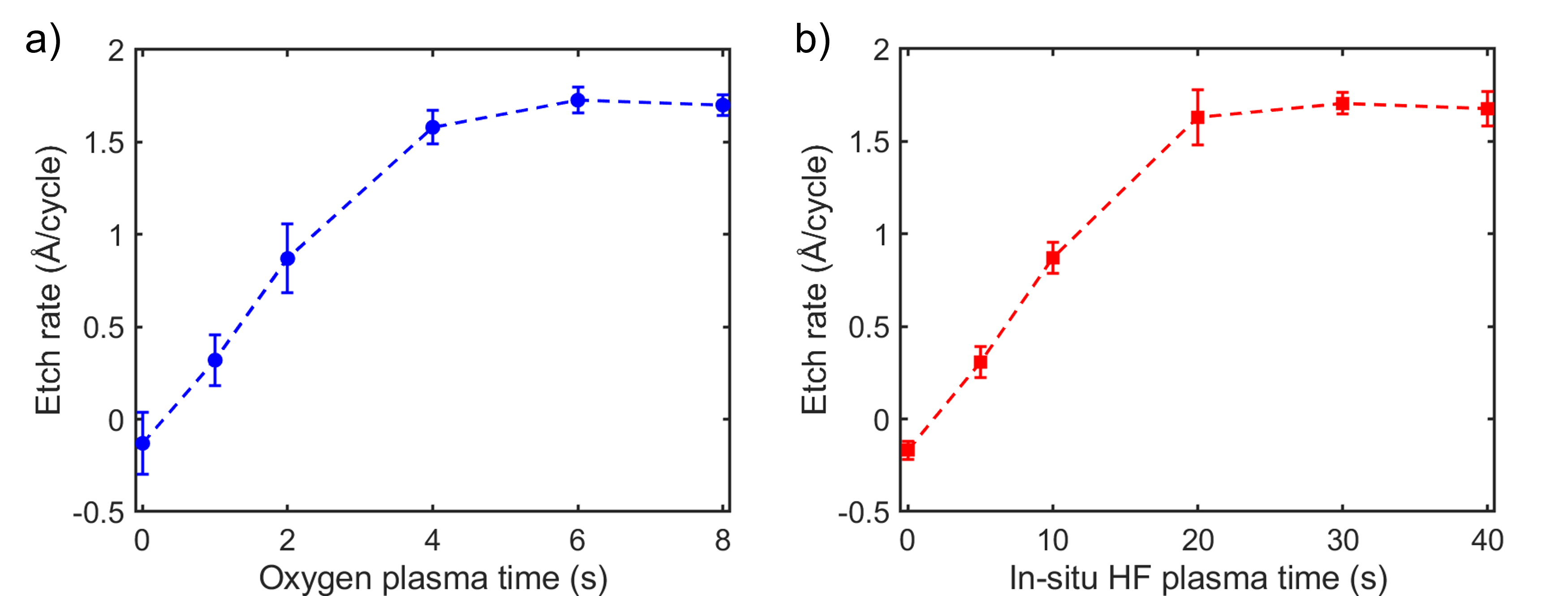}
\phantomsubcaption\label{fig:O2Sat}
\phantomsubcaption\label{fig:HFSat}
}
\caption{(a) NbN etch per cycle (EPC) versus O\sub2 plasma exposure time with \textit{in situ} HF exposure time fixed at 30 s. (b) EPC versus \textit{in situ} HF exposure time while O\sub2 plasma time is fixed at 6 s. The EPC is observed to saturate with increasing exposure time for each half step.}
\end{figure}

\Cref{fig:thicknesscycle} shows the thickness change of NbN versus number of cycles for the oxidation step, fluorination step, and for the full ALE recipe at \degC{125}. For the half-cycles, the thickness change was measured after exposure to only O\sub2 plasma or \textit{in situ} HF. No etching was observed for either half-cycle. In contrast, we observe a decrease in the thickness with increasing number of cycles when using both steps. The etch rate is calculated by dividing the total thickness change by the number of cycles, resulting in an etch per cycle (EPC) of  $1.77 \pm 0.06$ \Ang/cycle at \degC{125}. Based on the etch per cycle of the half cycles and overall cycle, the synergy is 100\% \cite{Kanarik2017Sep}. 

We also measured the saturation curves of each half step. \Cref{fig:O2Sat,,fig:HFSat} show EPC versus exposure time of O\sub2 plasma and \textit{in situ} HF, respectively. For each saturation curve the purge times and exposure time for one half-cycle is fixed while the other is changed. The reported EPC values were calculated from the thickness change measured after 30 cycles at \degC{125}. In \Cref{fig:O2Sat} the \textit{in situ} HF time is fixed at 30 s and the O\sub2 plasma time is varied. The etch rate is observed to saturate to $1.75 \pm 0.04$ \Ang/cycle after 6 s of O\sub2 plasma exposure. Similarly, in \Cref{fig:HFSat} the O\sub2 plasma time is fixed at 6 s while the \textit{in situ} HF exposure time is varied. The etch rate is found to saturate at $1.74 \pm 0.06$ \Ang/cycle after 30 s of \textit{in situ} HF exposure. The saturation of both half-cycles indicates that the process is atomic layer etching.

\subsection{Surface morphology characterization} \label{subsec:afm}

\begin{figure}
\centering

{\includegraphics[width = 0.9\textwidth]{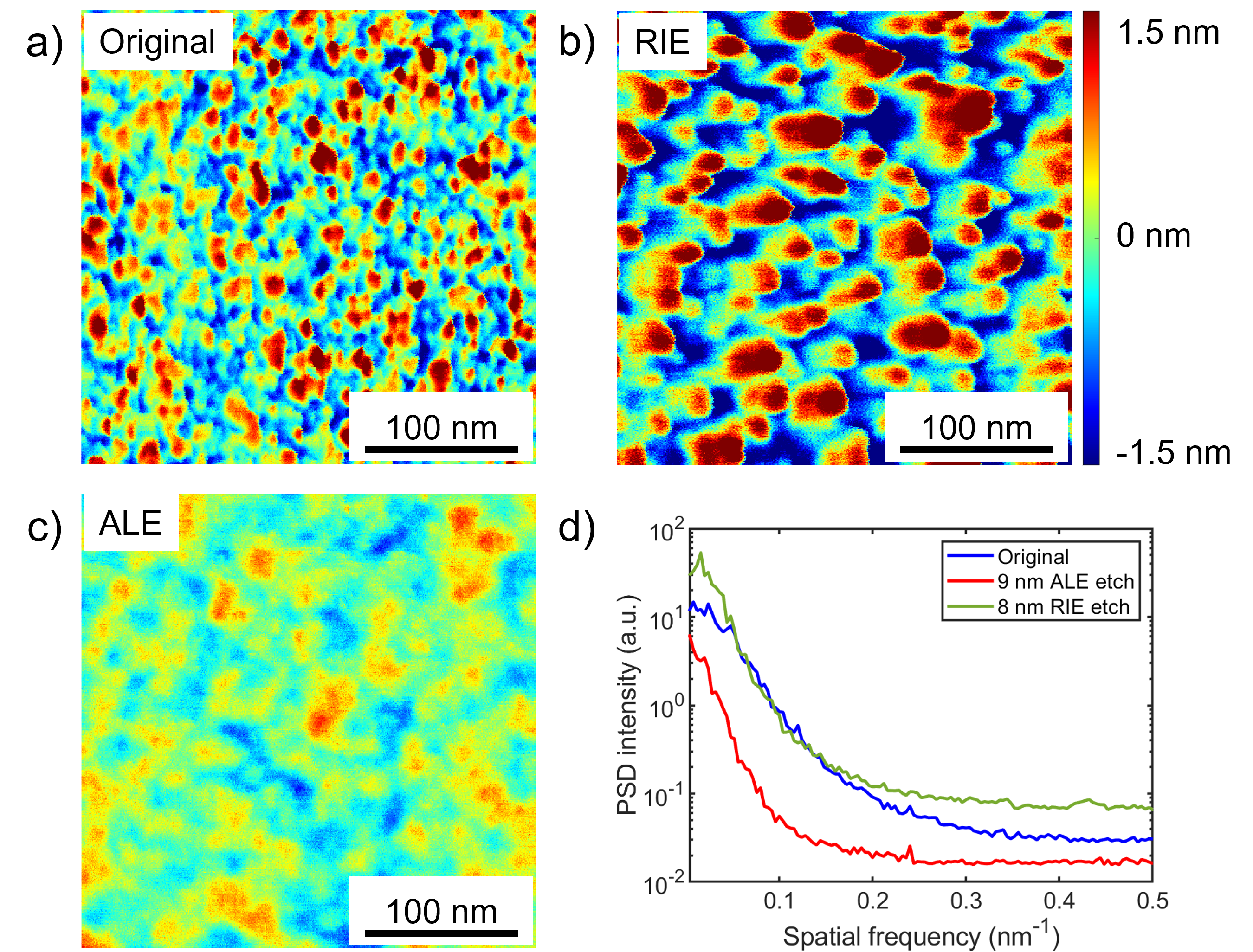}
\phantomsubcaption\label{fig:AFMOG}
\phantomsubcaption\label{fig:AFMRIE}
\phantomsubcaption\label{fig:AFMALE}
\phantomsubcaption\label{fig:PSD}}

\caption{AFM scan showing height-maps of (a) original sputtered sample, (b) after 8 nm etching with RIE, and (c) after 50 cycles ALE etching of the original sample, corresponding to  9 nm of etching. (d) PSD of samples (a)-(c) showing the roughening after RIE and smoothing after ALE.}
\label{fig:AFMHeight}

\end{figure}

We characterized the effect of ALE and RIE on surface morphology using AFM. \Cref{fig:AFMOG} shows a $250 \times 250$ nm$^2$ height-map of the original sputtered NbN film. The root mean square roughness of the scan was $R_q^{\text{original}} = 0.67$ nm. \Cref{fig:AFMRIE} shows a height-map of the RIE-treated film with $R_q^{\text{RIE}} = 0.97$ nm, which is 45\% rougher than the original. \Cref{fig:AFMALE} shows a height-map for the ALE-treated film after 50 cycles, which is 55\% smoother with $R_q^{\text{ALE}} = 0.30$ nm. The $R_q$ values were found to vary by $\lesssim 7\%$ between 3 locations on each film. 

\Cref{fig:PSD} shows the calculated PSD for the original, ALE-treated, and RIE-treated samples. A comparison of the PSDs at different spatial frequencies indicates the extent of smoothing or roughening at various length scales. The RIE film's PSD is within $20 \pm 5$\% of that of the original film over spatial frequencies from $\sim 0.05-0.14$ nm$^{-1}$. For spatial frequencies larger than $0.14$ nm$^{-1}$ and smaller than $0.05$ nm$^{-1}$, the RIE-treated surface is significantly rougher than the original sputtered surface, with the RIE PSD being more than $100\%$ larger at some spatial frequencies. We also observe that the ALE film's PSD is smaller than the original film's PSD across all spatial frequencies, indicating that ALE smooths the film over the length scales shown in \Cref{fig:PSD}.

\subsection{Surface and bulk film composition} \label{subsec:regxps}

\begin{figure}
\centering
{\includegraphics[width = 0.9\textwidth]{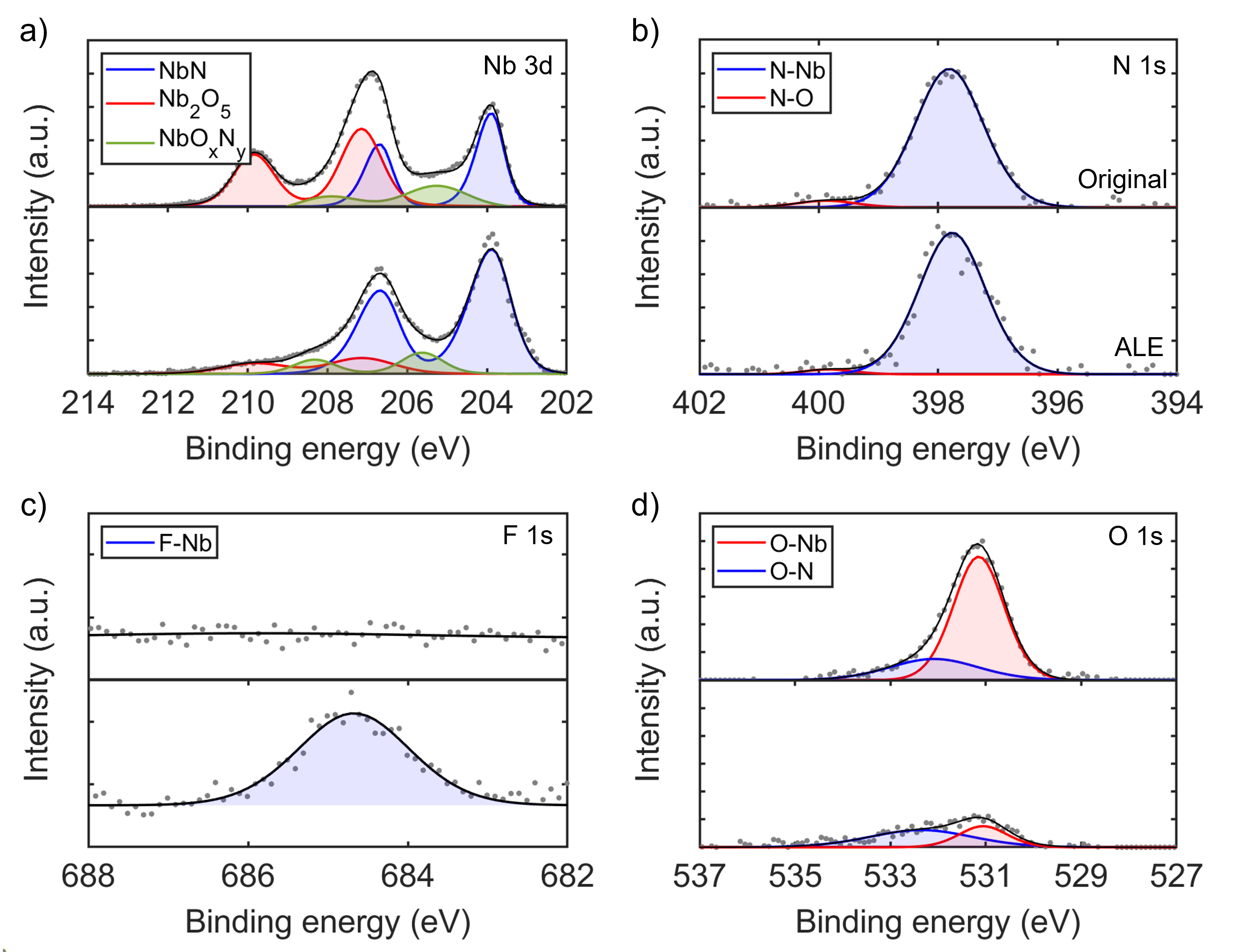}
\phantomsubcaption\label{fig:Nb3d}
\phantomsubcaption\label{fig:N1s}
\phantomsubcaption\label{fig:F1s}
\phantomsubcaption\label{fig:O1s}
}
\caption{Surface XPS spectra showing (a) Nb 3d, (b) N 1s, (c) F 1s, and (d) O 1s spectra. The spectra is shown for (top) original and (bottom) ALE-treated NbN films. The measured (gray dots) and fit spectra (black lines) intensity are reported in arbitrary units (a.u.) against the binding energy on the x-axis. The y-axis ticks are at the same intensity value for the original and ALE spectra within each element.}
\label{fig:CompXPS}
\end{figure}

We next characterize the surface composition of NbN before and after ALE. In \Cref{fig:CompXPS}, the XPS spectra for Nb 3d, N 1s, F 1s, and O 1s are shown for an untreated and an ALE-treated sample. In \Cref{fig:Nb3d}, we report the Nb 3d spectra for the original sample and ALE-treated film. We resolve three components consisting of doublets identical to the components shown in \Cref{fig:OxySurf}. We observe subpeaks corresponding to NbN (203.8 and 206.6 eV), NbO\sub{x}N\sub{y} (206.7 and 207.1 eV), and Nb\sub2O\sub5 (207.1 and 209.8 eV). The binding energy values corresponds to the 3d\sub{5/2} and 3d\sub{3/2} peaks, respectively. \Cref{fig:N1s} shows the N 1s spectra, where we observe peaks corresponding to N-Nb (397.6 eV) and N-O (399.6 eV) bonds. \Cref{fig:F1s} shows the F 1s spectra, where we observe one peak arising from Nb-F bonding (684.7 eV). In \Cref{fig:O1s}, we report the O 1s spectra with O-Nb (531.1 eV) and O-N (532.4 eV) subpeaks at corresponding to O-Nb and O-N bonding.

The top panels of \Cref{fig:CompXPS} are consistent with the XPS spectra of NbN thin films with a native oxide \cite{Yang2018May, Lubenchenko2018Jul, Darlinski1987Jun, Jouve1996Oct}. After ALE, we observe a decrease in the magnitude of O-Nb and Nb\sub{2}O\sub{5} components compared to the original spectra. This decrease in oxygen content after ALE has been observed in the ALE of other materials \cite{Wang2023May, Hennessy2017Jul, Metzler2017Jun, Hossain2023Sep}. A F-Nb peak appears in the F 1s spectra after ALE which was previously absent. The increase in fluorine after ALE using a fluorine-containing plasma has also been observed in many ALE recipes \cite{Fischer2017Mar, Catherall2024Sep, Wang2023May, Hennessy2017Jul, Chen2024Oct}.

\begin{figure}
\centering
{\includegraphics[width = \textwidth]{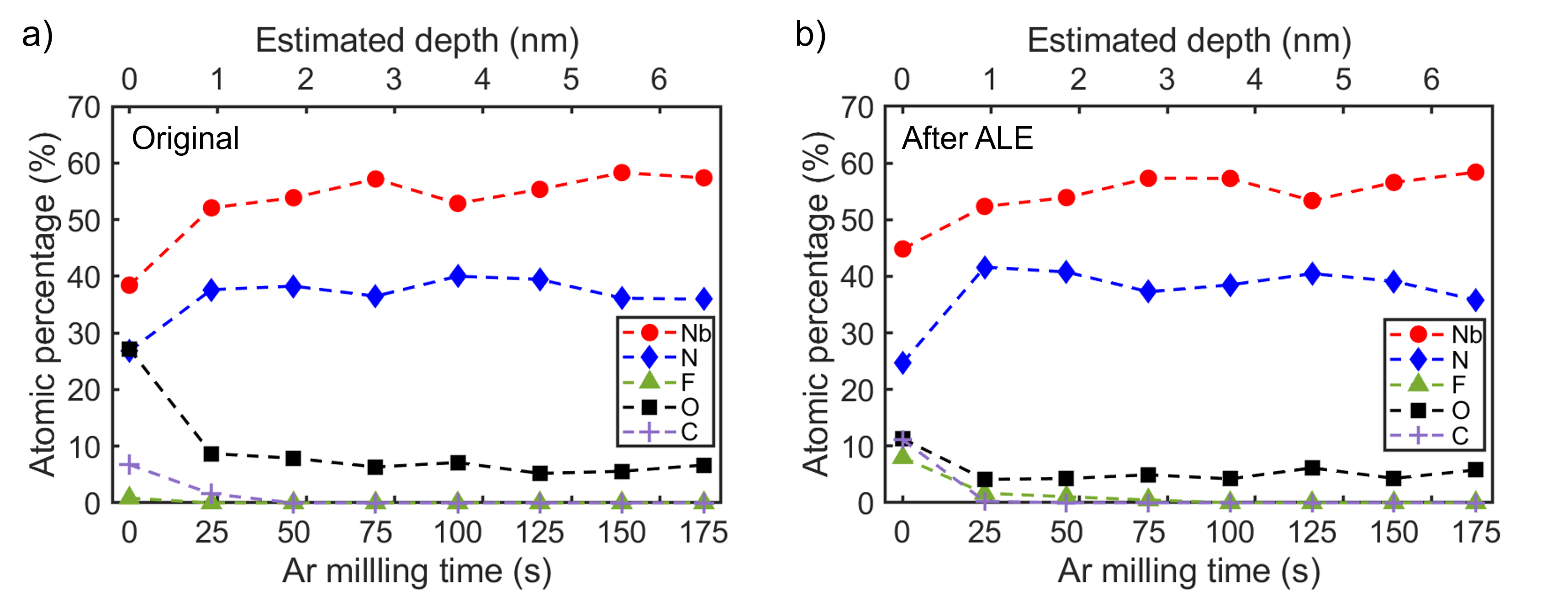}
\phantomsubcaption\label{fig:PreXPS}
\phantomsubcaption\label{fig:ALEXPS}
}
\caption{Atomic concentrations of Nb, N, F, O, and C versus Ar milling time and estimated depth for (a) original and (b) ALE-treated NbN thin films. The uncertainty for the data points is smaller than the size of the markers.}
\label{fig:depthXPS}
\end{figure}

We quantified the elemental composition at the surface and bulk of the original and ALE-treated films using depth-profile XPS. \Cref{fig:depthXPS} shows the atomic percentages of Nb, N, F, O, and C elements versus Ar milling time and estimated depth for the original sample and after 50 cycles of ALE. In \Cref{fig:PreXPS}, we report the surface composition of the original sample as 38.4\% Nb, 26.8\% N, 27.9\% O, 6.9\% C, and 0\% F. The composition of the original sample tends toward bulk values after 75 s of milling. The bulk atomic fractions are $\sim$ 56.2\% Nb, 37.6\% N, 6.2\% O, 0\% C, and 0\% F.

The surface and bulk values for Nb, N, and O are similar to those reported in other XPS studies of NbN \cite{Yang2018May, Lubenchenko2018Jul, Darlinski1987Jun, Jouve1996Oct, Cheng2019Dec}. The carbon on the surface is from C-O and C-H bonds typical of adventitious carbon caused by exposure to atmosphere. Oxygen contamination in bulk is often seen in sputtered polycrystalline Nb thin films \cite{Malev1995Oct, Manzo-Perez2024Sep}.

In \Cref{fig:ALEXPS}, we report the surface composition of the ALE-treated sample as 44.8\% Nb, 26.7\% N, 11.3\% O, 10.1\% C, and 7.1\% F. The composition of the ALE sample tends toward bulk values after 75 s of milling. The bulk atomic fractions are $\sim$ 56.3\% Nb, 37.9\% N, 5.8\% O, 0\% C, and 0\% F after 75 s milling or $\sim 3$ nm. After ALE, the surface oxygen concentration decreases by $59\%$. The surface fluorine is observed to increase after ALE which is consistent with prior work \cite{Fischer2017Mar, Catherall2024Sep, Wang2023May, Hennessy2017Jul, Chen2024Oct}. The fluorine contamination is observed to drop to below detection limits $(<0.1\%)$ after 50 s of Ar milling $(< 2 \text{ nm})$. The bulk atomic concentrations before and after ALE are found to vary by $<6\%$. Thus, we conclude that the effect of ALE is confined to the first few nanometers of the film and negligibly affects the bulk chemical composition.

\subsection{Cryogenic electrical properties} \label{subsec:resistivityandTc}

\begin{figure}
\centering
{\includegraphics[width = 0.7\textwidth]{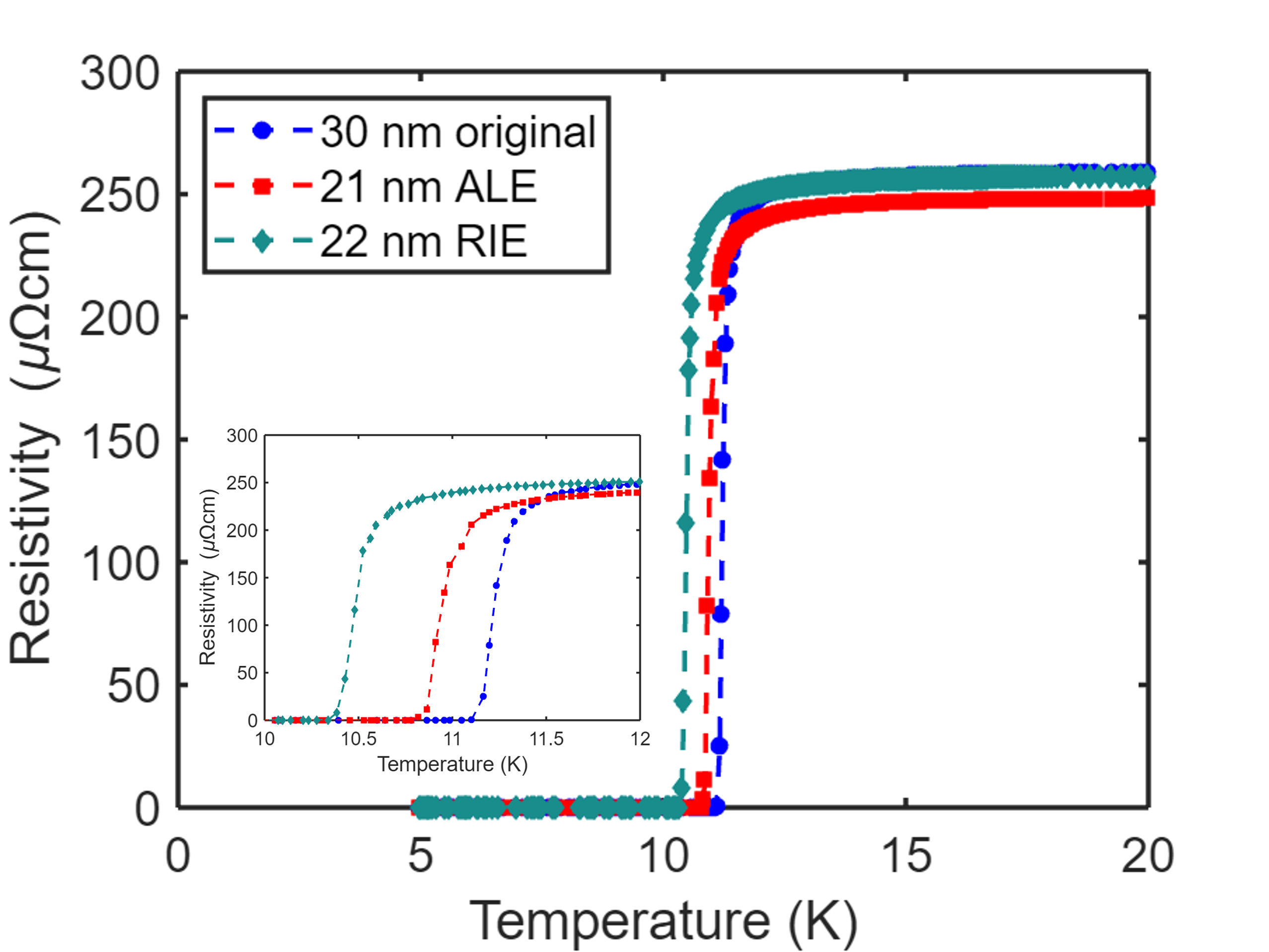}
}
\caption{Resistivity versus temperature for original 30 nm NbN film (blue circles), 21 nm thick ALE-treated film (red squares), and a 22 nm RIE-treated NbN film (green diamonds). Inset: Resistivity versus temperature plot with a truncated x-axis to highlight the superconducting transitions. The dashed lines are guides to the eye.}
\label{fig:PPMS}
\end{figure}

We investigated the effect of ALE and RIE on the electrical and superconducting properties of the NbN films by measuring their resistivity from 5 to 20 K. A 30 nm NbN film served as a reference. Another 30 nm film was etched using 50 cycles of ALE to 21 nm. Finally, a 30 nm NbN film was etched to 22 nm using RIE. \Cref{fig:PPMS} shows the measured resistivity versus temperature for the three films. The resistivity at 20 K and $T_c$ of the original 30 nm film were found to be $259\pm10\ \mu\Omega$cm and $11.21 \pm 0.07$ K, respectively. The resistivity at 20 K for the ALE-treated film was $249\pm9\ \mu\Omega$cm while for the RIE-treated film it was $257\pm10\ \mu\Omega$cm. We also measured $T_C^{\text{ALE}} = 10.95 \pm 0.07$ K and $T_C^{\text{RIE}} = 10.51 \pm 0.08$ K. While a decrease in $T_c$ with decreasing thickness is expected, we note that $T_c^{\text{ALE}} > T_c^{\text{RIE}}$ by 4.1\% even though the RIE-treated film is thicker than the ALE-treated film. We further estimate the $T_c$ of the ALE film if it were the same thickness as the RIE sample (22.1 nm) using a scaling law between film thickness and $T_c$ (see Eq.~1 in Ref.~\cite{Ivry2014Dec}). We calculate $T_c^{\text{ALE}} = 11.05$ K, corresponding to a 5.0\% higher $T_c$ compared to the RIE sample of the same thickness. We attribute this difference to the low damage nature of ALE compared to RIE. This experiment goes a step further than prior work \cite{Hossain2023Sep} by directly comparing the electrical and superconducting properties of samples treated by ALE and RIE, the latter being the state-of-the-art etching technology for superconducting devices at present.


\section{Discussion}

We now discuss the characteristics of our NbN ALE process in comparison to the closest analogous processes, ALE of TiN. ALE of TiN has been reported using two primary approaches: oxidation followed by fluorination to etch using volatile fluorides at the same temperature \cite{Lee2017TiN, Hossain2023Sep}, and fluorination or chlorination followed by thermal cycling to volatilize Ti-halides \cite{Miyoshi2022Apr, Shim2022}. The first approach achieves atomic-scale control with EPCs between 0.20 \Ang/cycle \cite{Lee2015Feb} and 2.4 \Ang/cycle \cite{Hossain2023Sep} at \degC{200}. However, Ref.~\cite{Lee2015Feb} uses HF vapor, which requires long purge times and is not routinely available in conventional processing tools. The second approach yields EPCs exceeding $17$ \Ang/cycle, which decreases processing times but lacks the etch depth precision desired for surface processing. Further, the use of thermal cycling can lead to impractical process times on conventional tools. 

The present NbN process achieves an EPC of 1.77 \Ang/cycle at \degC{125} and is most comparable to the TiN EPC of 2.4 \Ang/cycle at \degC{200} obtained in Ref.~\cite{Hossain2023Sep}, which also used an SF\sub{6}/H\sub{2} plasma instead of HF. However, the present recipe uses O\sub{2} plasma compared to the O\sub{2} gas used in Ref.~\cite{Hossain2023Sep}. The use of O\sub{2} plasma was necessitated by the complex chemistry of Nb and the need to maximize the proportion of the +5 oxidation state oxide. Our results indicate that O\sub{2} plasma increased the Nb\sub{2}O\sub{5} ratio by $\sim 250\%$ compared to O\sub{2} gas. The increase in Nb\sub{2}O\sub{5} content led to a reproducible etch per cycle over 50 cycles, which was not achievable using  O\sub{2} gas.


Our plasma-thermal ALE process may find potential applications in the etching of NbN-based superconducting quantum devices such as microwave kinetic inductance detectors, where the native oxide hosts parasitic TLS that presently limit the device performance. On the basis of our XPS and resistivity measurements, ALE-treated films have a reduced oxygen concentration while maintaining unaltered bulk chemistry and electrical properties. These properties make ALE promising for reducing the TLS density in the metal-air interface and thereby improving the quality factor of superconducting microresonators. The present process could be used as a post-treatment process by removing the few-nanometer-thick surface region hosting TLS after the primary etch process. The smoothing effect and Angstrom-scale EPC of the present process is also potentially relevant for fabricating NbN-based single nanowire single photon detectors where the sidewall roughness and other inhomogeneities negatively impact device metrics.

\section{Conclusion}

We have reported a plasma-thermal atomic layer etching process for NbN using sequential exposures of O\sub2 plasma and SF\sub{6}/H\sub{2} plasma. O\sub{2} plasma exposure yields a higher fraction of Nb\sub{2}O\sub{5} compared to O\sub2 gas exposure, enabling subsequent volatilization by SF\sub{6}/H\sub{2} plasma. The SF\sub{6}/H\sub{2} plasma selectively etches Nb\sub{2}O\sub{5} over NbN for suitable flow rate ratios. The etch rate is measured at 1.77 \Ang/cycle at \degC{125}. We observe a smoothing effect from ALE, corresponding to a $55\%$ reduction in RMS roughness after 50 cycles, along with a 59\% reduction in surface oxygen concentration. We also find that the $T_c$ of an ALE-treated sample is higher than that of RIE-treated sample of a similar thickness, highlighting the low-damage nature of the process. We anticipate that the ability to engineer the surface of NbN films on the Angstrom-scale will facilitate applications of NbN in superconducting quantum applications such as single-photon detectors and qubits.

\section{Acknowledgements}

This work was supported by NSF under Award \#2234390. The authors thank Emanuel Knehr and Sahil Patel (Jet Propulsion Laboratory) for supplying the NbN wafer used in this paper and Daniel N. Shanks (Jet Propulsion Laboratory) for useful discussions. We gratefully acknowledge the critical support and infrastructure provided for this work by The Kavli Nanoscience Institute and the Molecular Materials Research Center of the Beckman Institute at the California Institute of Technology.

\section{Data availability statement}

The data that support the findings of this study are available from the corresponding author upon reasonable request.

\section{Conflict of interest}

The authors have no conflicts to disclose.

\newpage
\clearpage

\bibliography{ref}

\end{document}